\documentclass[12pt,intlimits,reqno]{amsart}
\usepackage{enumerate}
\usepackage{amssymb}
\usepackage{amsthm}
\usepackage{graphics}
\usepackage{verbatim}
\usepackage{pstricks}
\usepackage{psfrag}
\newtheorem{thm}{Theorem}[section]

\theoremstyle{definition}

\newtheorem{ex}[thm]{Example}
\theoremstyle{remark}

\numberwithin{equation}{section}
\usepackage[cp1250]{inputenc}

\renewcommand{\sl}{sl(2,\R)}

\usepackage{enumerate}

\newcommand{\be}{\begin{equation}}
\newcommand{\ee}{\end{equation}}
\newcommand{\bse}{\begin{subequations}}
\newcommand{\ese}{\end{subequations}}

\renewcommand{\sc}[2]{\langle #1|#2 \rangle}

\newcommand{\proj}[2]{|#1 \rangle \langle #2|}

\newcommand{\ket}[1]{\left| #1 \right\rangle}
\newcommand{\abs}[1]{\left\vert#1\right\vert}

\newcommand{\R}{\mathbb R}

\newcommand{\No}{\mathbb{N}\cup\{0\}}

\newcommand{\Z}{\mathbb Z}
\newcommand{\C}{\mathbb C}
\newcommand{\N}{\mathbb N}
\renewcommand{\sl}{sl(2,\R)}
\renewcommand{\H}{\mathcal{H}}

\newcommand{\A}{\mathbf{A}}
\newcommand{\B}{\mathbf{B}}
\newcommand{\D}{\mathbf{D}}
\newcommand{\Dd}{\mathbb{D}}

\renewcommand{\a}{\mathbf{a}}
\newcommand{\Rr}{\mathbf{R}}
\newcommand{\Hh}{\mathbf{H}}
\newcommand{\n}{\mathbf n}
\newcommand{\tto}{\longrightarrow}
\newcommand{\K}{\mathcal{K}}
\renewcommand{\O}{\mathcal{O}}

\DeclareMathOperator{\spec}{spec}

\DeclareMathOperator{\spn}{span}
\DeclareMathOperator{\id}{id}
\DeclareMathOperator{\Aut}{Aut}

\begin{document}

\begin{center}
{\Large\bf  $\sl$ symmetry and solvable multiboson systems
}

\bigskip

{\bf Tomasz Goli\'nski\footnote{tomaszg@alpha.uwb.edu.pl}, Maciej Horowski\footnote{horowski@alpha.uwb.edu.pl}, Anatol Odzijewicz\footnote{aodzijew@uwb.edu.pl}, Aneta Sli\.zewska\footnote{anetasl@uwb.edu.pl}
}

\end{center}
\begin{center} {Institute of Mathematics\\
University in Bia{\l}ystok
\\Lipowa 41, 15-424 Bia{\l}ystok, Poland}
\end{center}
\bigskip\bigskip

\begin{abstract}
The one-mode and the two-mode multiboson systems with $\sl$ symmetry are investigated.
Hamiltonians of these systems are integrated using the theory of orthogonal polynomials. 
The coherent state representation for these systems is constructed.
\end{abstract}

\section*{Introduction}

We extend  the well known Holstein-Primakoff method to the case of $\sl$ algebra. The original method   
is based on investigation of quantum systems described by the elements of the enveloping algebra of Heisenberg-Weyl algebra satisfying
commutation relations of $su(2)$, see \cite{holstein}, and is most effective in the study of shell model of the nucleus. Various examples of that can be found in \cite{klein-marshalek}.
Our generalization is used to the integration of the quantum-optical models such as for example the model of the two-level atom coupled to single mode radiation, see e.g. \cite{sukumar-buck,sukumar}. 

In this paper we combine the $\sl$ symmetry and the theory of the orthogonal polynomials in order to integrate the wide family of multiboson Hamiltonians  describing the interaction of bosons with non-linear medium as well as the boson-boson interaction. Hence among other one can use them for modelling the non-linear phenomena in quantum optics including Kerr-like effect and parametric generation, see \cite{Peng-Li,HChOT}.

The paper consists of three sections. In the Section \ref{cha_1} we distinguish the two-parameter family of one-mode multiboson Hamiltonians \eqref{1-ham} defined as a linear combination of the generators of Lie algebra $\sl$ obtained by solving the  difference equations \eqref{f}-\eqref{pocz}. We define the group of the Bogoliubov-like transformations and using it we distinguish the classes of Hamiltonians related to the appropriate classes of orthogonal polynomials, see also Appendix \ref{ap-1-mode}.  At the end of the section, we construct the system of coherent states which allows us to find new, as we suppose, realization of the discrete series representation of the group $SL(2,\R)$.

Section \ref{cha_2} is devoted to solving the class of two-mode Hamiltonians \eqref{hamiltonian}, which are defined by the combination of Casimirs for $\sl$ algebras. Similarly to the one-mode case the integration of \eqref{hamiltonian} is carried out by the orthogonal polynomials, i.e by dual Hahn and continuous dual Hahn polynomials. 

In the last section we discuss the possible physical interpretation of the quantum models governed by the Hamiltonians investigated in previous sections.

\section{One-mode systems}\label{cha_1}
\subsection{Multiboson representations of $\sl$ in bosonic Fock space}$\;$\label{sec_prelim}

Let $\a$, $\a^*$  be the standard annihilation and creation operators satisfying
CCR relations and acting in the Hilbert space $\H$. 
Let $\{\ket{n}\}_{n=0}^\infty$ be the orthonormal basis in $\H$ consisting of eigenvectors of number occupation operator $\n:=\a^*\a$.

For any fixed $l\in\N$, we define the \emph{multiboson representation} of $\sl$ as the triple of operators
 \be \label{l-cluster}\A_0:=\alpha_0(\n), \qquad \A_-:=\alpha_-(\n)\;\a^{l},\qquad \A_+:=(\a^*)^l\alpha_-(\n)\ee
defined on dense subset $\mathcal D_0$ consisting of finite combinations of $\ket n$ and
satisfying $\sl$ commutation relations:
\be\label{SL_comm} [\A_-,\A_+]= \A_0, \qquad  [\A_0,\A_\pm]=\pm2 \A_\pm \ee
with symmetricity conditions:
\be\label{SL_symm}\A_0\subset\A_0^*,\qquad \A_-\subset\A_+^*, \qquad \A_+\subset\A_-^*.\ee
We denote by $\mathcal A$ the operator Lie algebra generated by $\A_0, \A_-, \A_+$.

Relations \eqref{SL_comm} and \eqref{SL_symm} imply that the functions $\alpha_0$, $\alpha_-$ are real-valued and satisfy the following difference equations
\begin{eqnarray}\label{f} \big(\alpha_0(n)-\alpha_0(n-l)-2\big)\alpha_-(n-l)=&0 &\textrm{for }n\geq l,\\
\label{alpha} (n+1)_l\;\alpha_-^2(n)-(n-l+1)_l\;\alpha_-^2(n-l)=&\alpha_0(n) &\textrm{for }n\geq l,\\
\label{pocz}(n+1)_l\;\alpha_-^2(n)=&\alpha_0(n) &\textrm{for }0\geq n< l,
\end{eqnarray}
where $(n)_l=n(n+1)\ldots(n+l-1)$ and  one has applied the identities
\be ({\a^*})^{l}\a^{l}=(\n-l+1)_l \qquad \a^{l}({\a^*})^l= (\n+1)_l.\ee

Without loss of generality we
can assume that $\alpha_-(n)> 0$ for $n=0,1,\ldots$ (by multiplying basis vectors by $\pm1$ and by removing common kernel of $\A_0$, $\A_-$, $\A_+$).

The solution to the system of difference equations \eqref{f}, \eqref{alpha}, \eqref{pocz} is of the form
\bse\label{sl-sol} \be\label{alpha-sol}\alpha_0(n)=2\left[\frac{n}{l}\right] + \alpha_0(n \bmod l),\ee
\be \label{phi-sol}\alpha_-(n)=\sqrt{\frac{1}{(n+1)_l}\left(\left[\frac{n}{l}\right]+\alpha_0(n\bmod l)\right)\left(\left[\frac{n}{l}\right]+1\right)},\ee\ese
where $[x]$ is the integer part of $x$. The values $\alpha_0(r)$ for $r=0,\ldots,l-1$ are arbitrary positive constants
corresponding to initial conditions on the solution $\alpha_0$ of the difference equation \eqref{f}.

In order to express the operators $\A_0$, $\A_-$, $\A_+$ explicitly in terms of the creation and annihilation operators let us define the bounded operator
\be \label{oper-R}\Rr:=\frac{l-1}2 + \sum_{m=1}^{l-1} \frac{\exp(-\frac{2\pi i m}{l} \n)}{\exp(\frac{2\pi i m}{l})-1}\ee
for $l>1$ and $ \Rr:=0$ for $l=1$, see \cite{HChOT}.
This operator acts on elements of the basis by
\be \Rr\ket{n}=n\bmod l \;\ket n\ee
and commutes with operators $\A_0$, $\A_-$, $\A_+$.
Thus one has
\be \frac1l \left(\n - \Rr\right)\ket n = \left[\frac nl \right]\ket n.\ee
Finally the multiboson representation of $\sl$ is given in terms of $\a$ and $\a^*$ by
\bse \label{sl-oper}\be \A_0=\frac2l \left(\n - \Rr\right)+\alpha_0(\Rr),\ee
\be \A_-=\sqrt{\frac1{(\n+1)_l}\left(\frac1l (\n - \Rr)+\alpha_0(\Rr)\right)\left(\frac1l (\n - \Rr)+1\right)}\;\a^l.\ee\ese
$\alpha_0$ is these formulae is treated as an arbitrary positive function on spectrum of $\Rr$, i.e. it is defined by initial conditions $\alpha_0(r)$,
$r\in\spec(\Rr)=\{0,1,\ldots l-1\}$, see \eqref{sl-sol}.

The formulae \eqref{sl-oper} show that the Hilbert space $\H$ splits
\be \label{dec-1mode}\H=\bigoplus_{r=0}^{l-1} \H_r\ee
onto invariant subspaces 
\be \label{sub-1mode}\H_{r}:=\spn\{ \;\ket{k}_r:=\ket{kl+r} \;|\; k\in\No \},\ee
which are eigenspaces of $\Rr$ corresponding to the eigenvalue $r$.

Let us observe that $\ket k_r$ are eigenvectors of $\A_0$
\be \label{A0k}\A_0\ket k_r=(2k+\alpha_0(r))\;\ket k_r\ee
and $\A_-$, $\A_+$ act on $\ket k_r$ as weighted shift operators:
\be \label{A-k}\A_-\ket k_r=\sqrt{k(k+\alpha_0(r)-1)}\;\ket {k-1}_r,\ee
\be \label{A+k}\A_+\ket k_r=\sqrt{(k+\alpha_0(r))(k+1)}\;\ket {k+1}_r.\ee
So, \eqref{dec-1mode} gives the decomposition of  the multiboson representation of $\sl$ onto irreducible components.

Let us now present two examples of the multiboson representation of $\sl$.

\begin{ex}\label{ex-hp}
  Let us begin by the simplest case $l=1$. Then $\Rr=0$ and
  $\alpha_0(\Rr)=c$. Therefore
\be \A_0=2\n +c, \qquad\qquad \A_-=\sqrt{\n + c}\;\a.\ee
It is $\sl$ analogue of Holstein-Primakoff mapping for $SU(2)$, see \cite{klein-marshalek}.\begin{flushright}$\Box$\end{flushright}
\end{ex}

\begin{ex}\label{ex-2}Let us put $l=2$
and the special choice of $\alpha_0$, namely $\alpha_0(\Rr)=\frac12+\Rr$.
Then formulae \eqref{sl-oper} simplify to:
\be \A_0=\n +\frac12,\qquad\qquad \A_-=\frac12\,\a^2.\ee
This representation is associated with the problem of second-harmonic generation, see \cite{walls-tindle,jex-drobny}.
\begin{flushright}$\Box$\end{flushright}\end{ex}

The Casimir operator for $\mathcal A$ is of the form
\be \label{casimir}\mathbf C_\mathcal A:=\frac12 \A_0^{\phantom02} -(\A_-\A_++\A_+\A_-)\ee
and  it can be explicitly expressed by
\be \mathbf C_\mathcal A=\frac12\alpha_0(\Rr)\left( \alpha_0(\Rr)-2\right).\ee
Thus $\mathbf C_\mathcal A|_{\H_r}=\frac12\alpha_0(r)\left( \alpha_0(r)-2\right)$ and each $\H_r$ is  contained in some eigenspace of $\mathbf C_\mathcal A$. For $0<\alpha_0(r)<2$
or $\alpha_0(r)=2,3,\ldots$ representation \eqref{A0k}-\eqref{A+k} corresponds respectively to the complementary and discrete series 
of unitary representations of the group $SL(2,\R)$, see~\cite{vilenkin}.

From  \eqref{A0k}
it follows that $\A_0$ is closable with closure defined on
\be \mathcal D_1:=\left\{ \sum_{n=0}^\infty v_n \ket n \in\H \;\;\bigg|\;\; \sum_{n=0}^\infty n^2\abs{v_n}^2<\infty\right\} .\ee
Equations \eqref{A-k}-\eqref{A+k} imply that $\A_-$ and $\A_+$ are superpositions of diagonal operator and shift operator.
Both of these diagonal operators are closable with closure defined on $\mathcal D_1$. Since shift operator is bounded we conclude that the closures of all generators of $\mathcal A$ are defined on common domain $\mathcal D_1$. From now on $\A_0$, $\A_-$ and $\A_+$ will denote these closures.  Moreover symmetricity conditions \eqref{SL_symm} are replaced by stronger conditions 
\be\label{SL_herm}\A_0=\A_0^*,\qquad \A_-=\A_+^*, \qquad \A_+=\A_-^*,\ee
i.e. operator $\A_0$ is self-adjoint and $\A_-$, $\A_+$ are mutually adjoint.  

\subsection{Bogoliubov-like transformations}\label{sec-bogo}$\;$

Let us consider a linear transformation of $\mathcal A$ preserving commutation relations \eqref{SL_comm} and  conditions \eqref{SL_herm}.
We will call such transformation a \emph{Bogoliubov transformation} for the algebra $\sl$ in the analogy to Bogoliubov transformations for Heisenberg-Weyl algebra, see \cite{bogoliubov}.

The group of all transformations of this type is  isomorphic to the group
$\mathfrak B:=\R^\times\rtimes\Z_2$, where $\R^\times:=\R\setminus\{0\}$, $\Z_2=\{-1,1\}$, with
the group operation defined by
\be (a,\sigma)\cdot(b,\tau):=(ab^{\sigma},\sigma\tau).\ee
One shows that the transformation given by an element  $(a,\sigma)\in\mathfrak B$ acts on the generators of $\mathcal A$ in the following way
\begin{eqnarray}
\label{bogo-A}
\mathfrak b_{a,\sigma}(\A_0)&:=& \frac{1+a^2}{2a}\A_0+\sigma\frac{1-a^2}{2a}(\A_-+\A_+),\\
\mathfrak b_{a,\sigma}(\A_-)&:=& \frac{1-a^2}{4a}\A_0+\sigma\frac{(1-a)^2}{4a}\A_++\sigma\frac{(1+ a)^2}{4a}\A_-\nonumber,\\
\mathfrak b_{a,\sigma}(\A_+)&:=& \frac{1-a^2}{4a}\A_0+\sigma\frac{(1+a)^2}{4a}\A_++\sigma\frac{(1- a)^2}{4a}\A_-\nonumber.
\end{eqnarray}

Let us notice that the domain of operators \eqref{bogo-A} is also $\mathcal D_1$ and the Casimir operator \eqref{casimir}
is invariant with respect to the action  \eqref{bogo-A}. 

There exists the unitary projective representation $(a,\sigma)\longmapsto\mathbb U_{a,\sigma}$ in $\H_r$ of the subgroup
$\R_+\rtimes\Z_2\subset\mathfrak B$ which implements the action of $\mathfrak b_{a,\sigma}$, i.e.
\be\label{implement} \mathfrak b_{a,\sigma} (\mathbf X) = \mathbb U_{a,\sigma}\mathbf X\,\mathbb U^*_{a,\sigma}\ee
for all $\mathbf X\in\mathcal A$ on $\mathcal D_1$.

In order to prove this we give the explicit formula for $\mathbb U_{a,\sigma}$. First of all 
let us observe that basis vectors $\ket k_r$ in $\H_r$ are eigenvectors of $\A_0$. We will show that
$\mathfrak b_{a,\sigma}(\A_0)$ has the same spectrum as $\A_0$ and we will compute its eigenvectors.
To this end we observe that 
\be\label{1-3-term}\mathfrak b_{a,\sigma}(\A_0)\ket k_r=b_{k-1}\;\ket{k-1}_r +
a_k\;\ket k_r+b_k\;\ket{k+1}_r,\ee
where
\bse\label{meixxxx}\be a_k=\frac{a^{-\sigma}+a^\sigma}2(2k+\alpha_0(r))\ee
\be b_k=\frac{a^{-\sigma}-a^\sigma}2\sqrt{(k+\alpha_0(r))(k+1)}.\ee\ese

If $a\neq1$ then the formula \eqref{1-3-term} is directly related to three term recurrence relation
\be \label{3term-poly}x P_k(x) = b_{k-1}\;P_{k-1}(x) + a_k\;P_k(x)+b_k\;P_{k+1}(x)\ee
which is valid for any orthonormal polynomials family $\{P_n\}_{n=0}^\infty$ for apropriate choice of coefficients. 

In case when coefficients $a_k$ and $b_k$ are given by \eqref{meixxxx} we obtain Meixner polynomials 
$P_n(x)=M_k(x;\alpha_0(r),c)$, where 
\be c= \left(\frac{a-1}{a+1}\right)^{2},\ee
see \cite{akhiezer-OPS,askey}. From this it follows that  
\be \label{meix-baza}\ket{n;a,\sigma}_r:=
                            \left\{ \begin{array}{ll}
\sigma^n\sqrt{\frac{n!}{(\alpha_0(r))_nc^n}}\sum_{k=0}\limits^\infty M_k(n;\alpha_0(r),c) \;\ket{k}_r  &\textrm{for }a^\sigma<1 \\
\sigma^n\sqrt{\frac{n!}{(\alpha_0(r))_nc^n}}\sum_{k=0}\limits^\infty (-1)^kM_k(n;\alpha_0(r),c) \;\ket{k}_r  &\textrm{for }a^\sigma>1 \\
\sigma^n\ket{n}_r \phantom{\sum_{k=0}\limits^\infty}&\textrm{for } a=1
                             \end{array}\right.\ee
and the corresponding eigenvalues are
\be \label{meix-ev}E_n^{(\sigma,a)}:= 2n+\alpha_0(r). \ee

Since the spectra of $\A_0$ and $\mathfrak b_{a,\sigma}(\A_0)$ coincide we define the unitary operator $\mathbb U_{a,\sigma}$ by
\be \label{U}\mathbb U_{a,\sigma}\ket n_r := \ket {n;a,\sigma}_r.\ee

Using the fact that $\mathfrak b_{a,\sigma}(\A_-)$ and $\mathfrak b_{a,\sigma}(\A_+)$ satisfy the commutation relations \eqref{SL_comm}, we can check that they act on eigenvectors  $\ket {n;a,\sigma}_r$ as weighted shift operators:
\be \mathfrak b_{a,\sigma}(\A_-)\ket {n;a,\sigma}_r=
\sqrt{n(n+\alpha_0(r)-1)}\;\ket {n-1;a,\sigma}_r ,\ee
\be \mathfrak b_{a,\sigma}(\A_+)\ket {n;a,\sigma}_r=
\sqrt{(n+\alpha_0(r))(n+1)}\;\ket {n+1;a,\sigma}_r .\ee

Thus the unitary operator $\mathbb U_{a,\sigma}$ defined by \eqref{U} satisfies \eqref{implement} on dense subset of $\H_r$ consisiting of finite linear combinations of vectors $\ket {n;b,\pi}_r$ for $n\in\No$ and $(b,\pi)\in\R_+\rtimes\Z_2$. In such way we have constructed the unitary representation of 
the group $\R_+\rtimes \Z_2$, which implements the action \eqref{bogo-A}. However, it is not possible to extend it to whole $\mathfrak B$ since $\mathfrak b_{-a,\sigma}$ can be decomposed into
\be \mathfrak b_{-a,\sigma}=\mathfrak b_{a,\sigma}\circ \mathfrak b_{-1,1},\ee
where 
\be\mathfrak b_{-1,1} (\mathbf X)=-\mathbf X\ee
for $\mathbf X\in\mathcal A$.

\subsection{Integrable one-mode Hamiltonians}$\;$\label{sec-1mode-ham}

The aim of this section is to integrate quantum system described by arbitrary self-adjoint operator
belonging to the multiboson algebra~$\mathcal A$:
\be \label{1-ham}\Hh_{\mu\nu}:=\frac{\mu+\nu}{2}\A_0+\frac{\mu-\nu}{2}(\A_-+\A_+),\ee
where $(\mu,\nu)\in\R^2\setminus\{(0,0)\}$.

Let us observe that
\be \label{3-term}\Hh_{\mu\nu} \ket k_r = b_{k-1}\;\ket{k-1}_r +
a_k\;\ket k_r+b_k\;\ket{k+1}_r,\ee
where
\bse\be a_k=\frac{\mu+\nu}{2}(2k+\alpha_0(r))\ee
\be b_k=\frac{\mu-\nu}{2}\sqrt{(k+\alpha_0(r))(k+1)}.\ee\ese
If $\mu\neq\nu$ the formula \eqref{3-term} is directly related to three term recurrence relation \eqref{3term-poly}
for some family of orthonormal polynomials $\{P_n\}_{n=0}^\infty$.
Since  $\sum\frac1{b_k}$ is divergent then 
there exists the unique measure $d\omega$ on $\R$ such the map $F$ given by
\be \label{iso-F}\H_r \ni \ket k_r \longmapsto F(\ket k_r):=P_k\in L^2(\R,d\omega)\ee
is the isomorphism of Hilbert spaces with the property that
\be \label{x}F\circ \Hh_{\mu\nu|\H_r} \circ F^{-1}=\hat x,\ee
where $\hat x$ is the operator of multiplication in $L^2(\R,d\omega)$,
see \cite{chihara,OHT}. Thus we gather that the spectrum $\spec(\Hh_{\mu\nu})$ 
is the support of measure $d\omega$. It means that by finding the measure $d\omega$ and polynomials $P_n$ 
we obtain the evolution flow
\be R\ni t\tto e^{it \Hh_{\mu\nu}}=F^{-1}\circ e^{it\hat x}\circ F\in \Aut(\H_r)\ee
of quantum system described by the Hamiltonian $\Hh_{\mu\nu}$.

From definition of Bogoliubov group ${\mathfrak B}$ it follows that 
transformations ${\mathfrak b_{a,\sigma}}$ preserve the
family of operators $\Hh_{\mu\nu}$ and  
the labels $(\mu,\nu)$ transform as follows
\be \label{bogo-ham}(a,1):(\mu,\nu)\mapsto(a^{-1}\mu,a\nu),\qquad (a,-1):(\mu,\nu)\mapsto (a\nu,a^{-1}\mu). \ee
This defines the action of the group ${\mathfrak B}$ in the set $\R^2\setminus\{(0,0)\}$ of labels.
We conclude that 
orbits of $\mathfrak B$ are pairs of hiperbolae indexed by one real parameter $c\in\R$
\be \label{orbity}\O_c:= \mathfrak B \cdot (c,1)= \{ (x,y)\in\R^2\setminus\{(0,0)\} \;|\; xy=c\}.\ee

We can restrict our considerations to each component $\H_r$ of decomposition \eqref{dec-1mode} separately since they are invariant
under the action of $\Hh_{\mu\nu}$. Due to the implementation formula \eqref{implement} it is sufficient to find spectral decomposition for one operator from each orbit \eqref{orbity}, e.g. $\Hh_{\sqrt c,\sqrt c}$, $\Hh_{\sqrt c,-\sqrt c}$ and $\Hh_{1,0}$. Taking into account scaling by constant we can further restrict ourselves to three Hamiltonians $\Hh_{1,1}$, $\Hh_{1,-1}$ and $\Hh_{1,0}$.

Since $\Hh_{1,1}= \A_0$ then the spectral problem is trivial and is solved by formula \eqref{A0k}.

For $\Hh_{1,0}$ formula \eqref{3term-poly} defines Laguerre orthonormal polynomials 
\be P_n(x)=L_n^{(\alpha_0(r)-1)}(2x)\ee
and measure $d\omega$ is given by
\be d\omega(x)=2 (2x)^{\alpha_0(r)-1} e^{-2x} \theta(x) dx,\ee
where $\theta$ is Heaviside function.
Thus the spectrum $\spec(\Hh_{1,0})=\R_+\cup\{0\}$.

For $\Hh_{1,-1}$ formula \eqref{3term-poly} defines  Meixner-Pollaczek orthonormal polynomials 
\be P_n(x)=P_n^{(\frac{\alpha_0(r)}2)}\left(\frac{x}{2};\frac\pi 2\right)\ee
and measure is $d\omega$ is given by
\be d\omega(x)=\frac{1}{2}\abs{\Gamma\left(\frac12\alpha_0(r)+i\frac{x}{2}\right)}^2dx\ee
and the spectrum $\spec(\Hh_{1,-1})=\R$, see \cite{askey}.

\begin{figure}[h]\begin{center}
\includegraphics{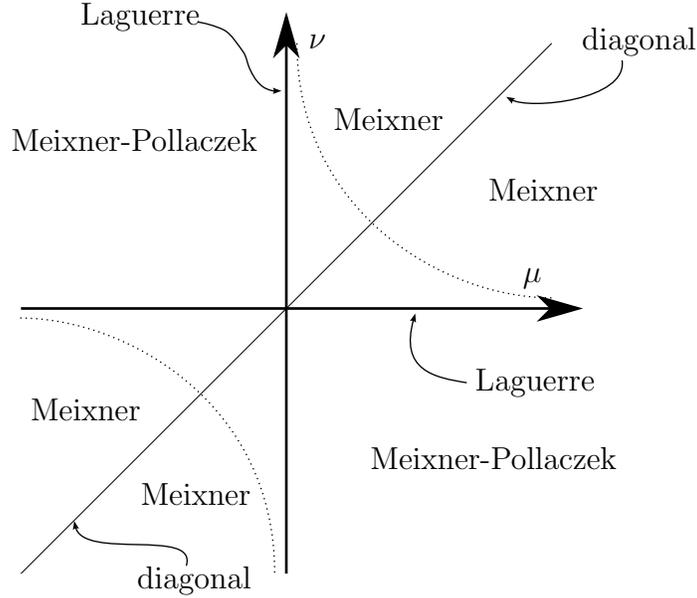}
\caption{Orthogonal polynomials assigned to $\H_{\mu\nu}$}
\label{fig-munuplane}
\end{center}\end{figure}

In Appendix \ref{ap-1-mode} we present complete list (for all $\mu$, $\nu$) of  Hilbert spaces $L^2(\R,d\omega)$ in which $\Hh_{\mu\nu}$ act as $\hat x$.
In Figure \ref{fig-munuplane} we have illustrated the decomposition of $(\mu,\nu)$-plane into sectors 
corresponding to different families of orthogonal polynomials. Let us remark that in the first quadrant the spectra of $\Hh_{\mu\nu}$ 
are discrete and bounded from below. On the boundary of this quadrant the spectra are $\R_+\cup\{0\}$. In the third quadrant
all spectra are also discrete but bounded from above. On the boundary of this quadrant the spectra are $\R_-\cup\{0\}$.
In the second and fourth quadrants the spectra are $\R$.

\psfrag{mu}{$\mu$}
\psfrag{nu}{$\nu$}
\psfrag{diagonal}{diagonal}
\psfrag{Meixner}{Meixner}
\psfrag{Meixner-Pollaczek}{Meixner-Pollaczek}
\psfrag{Laguerre}{Laguerre}

\subsection{Coherent state representation}\label{sec_coh}$\;$

The goal of this section is to give the coherent state representation
of the Hamiltonians $\Hh_{\mu\nu}$ and flows $e^{it\Hh_{\mu\nu}}$ generated by them.

We will consider coherent states as eigenstates of $\A_-$, see \cite{Oqspec}. Due to the decomposition \eqref{dec-1mode}
into irreducible representations, it is sufficient to restrict our considerations to each $\H_r$ separately
\be \A_- \ket\zeta_r = \zeta\ket\zeta_r.\ee
From \eqref{A-k} we see that the coherent states $\ket\zeta_r\in\H_r$ are given by the series
\be \ket\zeta_r:=\sum_{k=0}^\infty\frac{\zeta^k}{\sqrt{k!(\alpha_0(r))_{k} }} \ket k_r\ee
which converges for any $\zeta\in\C$ and belongs to domain $\mathcal D_1$.

The notion of the coherent states allows us to construct the anti-unitary embedding
\be \label{I}\H_r\ni\ket\psi\longmapsto I_r(\psi)(\zeta):=\sc\psi\zeta_r\in L^2\mathcal O(\C,d\mu_r)\ee
of $\H_r$ into the
Hilbert space $L^2\mathcal O(\C,d\mu_r)$ of holomorphic functions on $\C$, which are square integrable with respect to the measure
\be d\mu_r(\zeta,\overline\zeta):=\frac{\rho^{\alpha_0(r)} K_{\alpha_0(r)}(2\rho)}{2\pi \Gamma(\alpha_0(r))} \rho \;d\rho \;d\phi,\ee
where $\zeta=\rho e^{i\phi}$ and $K_{\alpha_0(r)}$ is the modified Bessel function of the second kind, see \cite{OS}. 
The space $L^2\mathcal O(\C,d\mu_r)$ has the reproducing kernel
\be \K(\overline\eta,\zeta):=\sc\eta\zeta=\;_0F_1\left(\begin{array}{@{}c|l@{}}-\\ \alpha_0(r) &\end{array}\overline \eta\zeta\right),\ee
i.e. for any $f\in L^2\mathcal O(\C,d\mu_r)$ one has
\be \int_\C \K(\overline\eta,\zeta) f(\eta)d\mu_r(\eta,\overline\eta)= f(\zeta).\ee
The isomorphism $I_r$ gives the realization of the operators $\A_0$, $\A_+$, $\A_-$ as the differential operators
\begin{eqnarray} 
\label{holo-A0}I_r\circ \A_0 \circ I_r^{\phantom r-1}&=&2\zeta \frac{d}{d\zeta}+\alpha_0(r),\\
 I_r\circ \A_+ \circ I_r^{\phantom r-1}&=&\zeta,\\
\label{holo-A-} I_r\circ \A_- \circ I_r^{\phantom r-1}&=&\left(\alpha_0(r)+\zeta \frac{d}{d\zeta}\right)\frac{d}{d\zeta}\end{eqnarray}
acting in $L^2\mathcal O(\C,d\mu_r)$. 
In order to describe them as the generators of the discrete series $\alpha_0(r)=2,3,\ldots$ representation of the group $SL(2,\R)$, 
let us consider a unitary integral transform
\be \mathcal P:L^2\mathcal O(\C,d\mu_r)\tto L^2\mathcal O(\Dd,d\nu_r),\ee
where $\Dd:=\{z\in\C\;|\; \abs z<1\}$ and
\be d\nu_r(z,\overline z):=\frac{\alpha_0(r)-1}{\pi}\left(1-\abs z^2\right)^{\alpha_0(r)-2}\;d^{\,2}\!z,\ee
given by
\be \label{perelomov}\mathcal Pf(z):=\int_\C e^{z \overline\zeta} f(\zeta) d\mu_r(z,\overline z).\ee

Using \eqref{perelomov} we find that in the space $L^2\mathcal O(\mathbb D,d\nu_r)$ the operators \eqref{holo-A0}-\eqref{holo-A-} are given by
\begin{eqnarray}
\mathcal P\circ I_r\circ\A_0 \circ I_r^{-1}\circ \mathcal P^{-1}&=&2z \frac{d}{dz}+\alpha_0(r)\nonumber\\
\mathcal P\circ I_r\circ\A_+ \circ I_r^{-1}\circ \mathcal P^{-1}&=& z^2\frac{d}{dz}+\alpha_0(r)z\\
\mathcal P\circ I_r\circ\A_- \circ I_r^{-1}\circ \mathcal P^{-1}&=&\frac{d}{dz}\nonumber
\end{eqnarray}
and they are the generators of the discrete series representation
\be \label{su-repr}U^{\alpha_0(r)}_g\phi(z)=(bz+\overline a)^{-\alpha_0(r)} \phi\left(\frac{az+\overline b}{bz+\overline a}\right)\ee
of the group $SL(2,\R)$ in $L^2\mathcal O(\mathbb D,d\nu_r)$, see \cite{perelomov}. Here we have identified $SL(2,\R)$ with $SU(1,1)$ using the isomorphism
\be SL(2,\R)\ni g \longleftrightarrow\left(\begin{array}{@{}cc@{}}
 a & b \\
 \overline b & \overline a \\
\end{array}\right):=
\frac12\left(\begin{array}{@{}cc@{}}
  1 & -i \\
  -i & 1 \\
\end{array}\right)
g
\left(\begin{array}{@{}cc@{}}
  1 & i \\
  i & 1 \\
\end{array}\right)\in SU(1,1).\ee

From \eqref{su-repr}  we obtain the representation of  $SL(2,\R)$ in $L^2\mathcal O(\C,d\mu_r)$ in the integral form
\be \label{group-action}T_g^{\alpha_0(r)}f(\zeta):=\int_\C \mathfrak K_g^{\alpha_0(r)}(\zeta,\overline\eta)\;f(\eta)\; d\mu_r(\eta,\overline\eta)\ee
with the kernel
\be \mathfrak K_g^{\alpha_0(r)}(\zeta,\overline\eta):=\int_\Dd e^{w\overline\eta-\zeta\frac{a\overline w - b}{\overline b\overline w-\overline a}}\;
\left(-\overline b\overline w+\overline a\right)^{\alpha_0(r)} \;d\nu_r(w,\overline w).\ee

In order to obtain the explicit formula for the evolution operators $e^{i\Hh_{\mu\nu}t}$
let us consider the corresponding family of one-parameter subgroups of $SL(2,\R)$
\be g_{\mu\nu}(t):=\cos\left(\sqrt{\mu\nu}\;t\right)\left(\begin{array}{@{}cc@{}}
  1& 0 \\
  0 & 1 \\
\end{array}\right)+\frac{\sin\left(\sqrt{\mu\nu}\;t\right)}{2\sqrt{\mu\nu}}
\left(\begin{array}{@{}cc@{}}
  -\mu+\nu& \mu+\nu \\
  -\mu-\nu & \mu-\nu \\
\end{array}\right)\ee

The infinitesimal generator of the action $T_{g_{\mu\nu}(t)}^{\alpha_0(r)}$
on $L^2\mathcal O(\C,d\mu_r)$ is $I_r\circ (i\Hh_{\mu\nu}t) \circ I_r^{\phantom r-1}$.
Thus the time evolution operator $I_r\circ (e^{i\Hh_{\mu\nu}t}) \circ I_r^{\phantom r-1}$ acts in $L^2\mathcal O(\C,d\mu_r)$ as the integral operator \eqref{group-action}
with the kernel
\be \label{evol-kernel}\mathfrak K_{g_{\mu\nu}(t)}^{\alpha_0(r)}(\zeta,\overline\eta):=\int_\Dd e^{w\overline\eta-\zeta\frac{a_{\mu\nu}(t)\overline w - b_{\mu\nu}(t)}{\overline{b_{\mu\nu}(t)}\overline w-\overline{a_{\mu\nu}(t)}}}\;
\left(-\overline{b_{\mu\nu}(t)}\overline w+\overline{a_{\mu\nu}(t)}\right)^{\alpha_0(r)} d\nu_r(w,\overline w),\ee
where
\be \label{hom-a}a_{\mu\nu}(t)=\cos\left(\sqrt{\mu\nu}\;t\right)+\frac{i\sin\left(\sqrt{\mu\nu}\;t\right)(\mu+\nu)}{2\sqrt{\mu\nu}}\ee
\be \label{hom-b}b_{\mu\nu}(t)=\frac{i\sin\left(\sqrt{\mu\nu}\;t\right)(-\mu+\nu)}{2\sqrt{\mu\nu}}.\ee

Ending let us remark that in $L^2\mathcal O(\C,d\mu_r)$ 
the Hamiltonian $\Hh_{\mu\nu}$ is represented as a second order differential operator
\be I_r\circ \Hh_{\mu\nu} \circ I_r^{-1} = \frac{\mu+\nu}{2}\left(2\zeta \frac{d}{d\zeta}+\alpha_0(r)\right)+\frac{\mu-\nu}{2}
\left(\zeta+\left(\alpha_0(r)+\zeta \frac{d}{d\zeta}\right)\frac{d}{d\zeta}\right)\ee
and in $L^2\mathcal O(\mathbb D,d\nu_r)$ as a  first order  differential operator
\be \mathcal P \circ I_r\circ \Hh_{\mu\nu} \circ I_r^{-1} \circ \mathcal P^{-1}=\frac{\mu+\nu}{2}\left(2z \frac{d}{dz}+\alpha_0(r)\right)+\frac{\mu-\nu}{2}\left((z^2+1)\frac{d}{dz}+\alpha_0(r)z\right).\ee
In Section  \ref{sec-1mode-ham} it was shown that this operator has discrete spectrum for $\mu\nu>0$ 
The eigenproblem 
\be I_r\circ \Hh_{\mu\nu} \circ I_r^{-1}\psi=\lambda\psi\ee
in the Hilbert space $L^2\mathcal O(\C,d\mu_r)$ 
leads to the degenerate Gauss equation and its solutions can be expressed in terms of confluent hypergeometric function and Laguerre-type function, see \cite{smirnov}.
On the other hand the eigenproblem  
\be \mathcal P\circ I_r\circ \Hh_{\mu\nu} \circ I_r^{-1}\circ \mathcal P^{-1}\phi=\lambda\phi\ee
in the Hilbert space $L^2\mathcal O(\mathbb D ,d\nu_r)$ leads to the first order differential equation and its solution in the case $\mu>\nu>0$  for the eigenvalue $\lambda$  is given by
\be \phi(z)=C \left(z-+\frac{(\sqrt \mu -\sqrt\nu)^2}{\mu-\nu}\right)^A \left(z+\frac{(\sqrt \mu +\sqrt\nu)^2}{\mu-\nu}\right)^B,\ee
where 
\be A=\frac {\lambda (\mu-\nu)}{2\sqrt{\mu\nu}}-\frac{\alpha_0(r)}{2}\qquad B=\frac {-\lambda (\mu-\nu)}{2\sqrt{\mu\nu}}-\frac{\alpha_0(r)}{2}.\ee
Thus 
\be \psi= (\mathcal P\circ I_r)^{-1} \phi\ee
and this kind of relationship gives the  
various integral representations of corresponding special functions. 
However the investigation of these problems is not the goal of the paper.

\section{Two-mode systems}\label{cha_2}

\subsection{Direct sum of multiboson $\sl$ algebras}$\;$\label{sec_dirsum}

In this section we will work within two-mode bosonic Fock space $\H\otimes\H$
with the standard annihilation, creation operators $\a_i, \a_i^*$, $i=0,1$, acting in $i^{\rm th}$ mode and
with the orthonormal basis $\{\ket{n_0,n_1}\}_{n_0,n_1=0}^\infty$ consisting of eigenvectors of $\n_i=\a^*_i\a_i$.

Let us consider the two-mode analogue of the investigation presented in the previous section. Let $\mathcal A\oplus \mathcal B\cong so(2,2)$ be the 
direct sum of two multiboson $\sl$ operator algebras  
$\mathcal A:=\spn\{\A_0,\A_-,\A_+\}$ and $\mathcal B:=\spn\{\B_0,\B_-,\B_+\}$ acting on $\mathcal D_0 \otimes \mathcal D_0\subset\H\otimes\H$, where
\be \label{2mode-A}\A_0:=\alpha_0(\n_0,\n_1) \qquad \A_-:=\alpha_-(\n_0,\n_1)\;\a_0^{\phantom 0 l_0}\qquad \A_+:=(\a_0^*)^{l_0}\alpha_-(\n_0,\n_1)\;\ee
\be \label{2mode-B}\B_0:=\beta_0(\n_0,\n_1) \qquad \B_-:=\beta_-(\n_0,\n_1)\;\a_1^{\phantom 1 l_1}\qquad \B_+:=(\a_1^*)^{	l_1}\beta_-(\n_0,\n_1)\ee
and $\mathcal A$ and $\mathcal B$ satisfy \eqref{SL_comm} and \eqref{SL_symm}

Proceeding as in previous section we gather that the commutation relations for $\mathcal A\oplus\mathcal B$
imply the
difference equations on functions $\alpha_0$, $\alpha_-$, $\beta_0$, $\beta_-$. The solution  of these equation gives the following expressions on operators:
\bse \label{sl-operA}\be \A_0=\frac2{l_0} \left(\n_0 - \Rr_0\right)+\alpha_0(\Rr_0,\Rr_1)\ee
\be \A_-=\sqrt{\frac1{(\n_0+1)_{l_0}}\left(\frac1{l_0} (\n_0 - \Rr_0)+\alpha_0(\Rr_0,\Rr_1)\right)\left(\frac1{l_0} (\n_0 - \Rr_0)+1\right)}\;\a_0^{\phantom 0 l_0},\ee\ese
\bse \label{sl-operB}\be \B_0=\frac2{l_1} \left(\n_1 - \Rr_1\right)+\beta_0(\Rr_0,\Rr_1)\ee
\be \B_-=\sqrt{\frac1{(\n_1+1)_{l_1}}\left(\frac1{l_1} (\n_1 - \Rr_1)+\beta_0(\Rr_0,\Rr_1)\right)\left(\frac1{l_1} (\n_1 - \Rr_1)+1\right)}\;\a_1^{\phantom 1 l_1},\ee\ese
where $\Rr_i$, $i=0,1$, are defined by \eqref{oper-R} for $i^{\rm th}$ mode
and $\alpha_0$, $\beta_0$ are arbitrary positive functions on the product of spectra $(\spec\Rr_0)\times(\spec\Rr_1)$.

We observe that $\Rr_i$ commute with $\mathcal A\oplus\mathcal B$
and that Casimir operators
$\mathbf C_{\mathcal A}$, $\mathbf C_{\mathcal B}$ (see \eqref{casimir})
can be expressed as functions (not invertible in general) of
$\Rr_0$ and $\Rr_1$. Thus
we will decompose two-mode Fock space 
\be \H\otimes\H=\bigoplus_{r_0=0}^{l_0-1}\bigoplus_{r_1=0}^{l_1-1}\H_{r_0,r_1},\ee
into common eigenspaces of $\Rr_i$
\be \H_{r_0,r_1}:=\spn\{ \;\ket{k_0,k_1}_{r_0,r_1}:=\ket{k_0l_0+r_0,k_1l_1+r_1} \;|\; k_0,k_1\in\No \},\ee
on which $\mathcal A\oplus\mathcal B|_{\H_{r_0,r_1}}$ acts irreducibly.

Comparing the formulae \eqref{sl-operA}-\eqref{sl-operB} to \eqref{sl-oper} we extend all generators of $\mathcal A\oplus\mathcal B$ to the common domain $\mathcal D_1\otimes \mathcal D_1$.

\subsection{Hamiltonians}$\;$

Let us consider a class of $\sl$ subalgebras $\mathcal D:=\spn\{\D_0,\D_-,\D_+\}\subset\mathcal A\oplus \mathcal B$ defined by 
\begin{eqnarray} \label{bogoD}\D_0&:=&\mathfrak b_{a,\sigma}(\A_0)+\mathfrak b_{b,\tau}(\B_0),\\
\D_-&:=&\mathfrak b_{a,\sigma}(\A_-)+\mathfrak b_{b,\tau}(\B_-),\nonumber\\
\D_+&:=&\mathfrak b_{a,\sigma}(\A_+)+\mathfrak b_{b,\tau}(\B_+),\nonumber\end{eqnarray}
for $(a,\sigma), (b,\tau)\in \mathfrak B$. For simplicity we have omitted indices in $\D_0$, $\D_-$, $\D_+$
corresponding to the elements of $\mathfrak B$.

Since $\D_0$, $\D_-$, $\D_+$ satisfy \eqref{SL_comm}, the Casimir for this subalgebra is of the form
\be \mathbf C_{\mathcal D}=\frac12 \D_0^{\phantom02}-\D_-\D_+-\D_+\D_-\ee
and it commutes with the Casimirs $\mathbf C_{\mathcal A}$ and $\mathbf C_{\mathcal B}$.

Main goal of this section is to integrate the two-mode bosonic quantum system described by the Hamiltonian
\be \label{hamiltonian}\Hh:=\frac12\left(\mathbf C_{\mathcal D}-\mathbf C_{\mathcal A}-\mathbf C_{\mathcal B}\right).\ee
Casimir operator $\mathbf C_{\mathcal D}$ is not defined on $\mathcal D_1\otimes\mathcal D_1$ but due to the 
explicit formula 
\begin{eqnarray} \label{ham-AB}\Hh&=& \frac{(a^2+b^2)}{4ab}\A_0\B_0-\\
&-&\sigma\tau\frac{(a-b)^2}{4ab}(\A_+\B_++\A_-\B_-)-\sigma\frac{a^2-b^2}{4ab}(\A_+\B_0+\A_-\B_0)+\nonumber\\
&+&\tau\frac{a^2-b^2}{4ab}(\A_0\B_-+\A_0\B_+)
-\sigma\tau\frac{(a+b)^2}{4ab}(\A_+\B_-+\A_-\B_+)\nonumber\end{eqnarray}
we see that the symmetric operator $\Hh$ is well defined on $\mathcal D_1\otimes\mathcal D_1$. 

Besides obvious integral of motions $\Rr_0$, $\Rr_1$ this Hamiltonian also commutes with $\D_0$ which is so called Manley-Rowe integral of motion for $\Hh$.
It means that the investigation of the 
the spectral decomposition of $\Hh$ can be reduced to the eigenspaces of $\D_0$ in each $\H_{r_0,r_1}$.

Due to the implementation formula \eqref{implement} we can unitarily transform $\Hh$ by 
$\mathbb U_{\abs a^{\sigma},-\sigma}\otimes \id + \id\otimes \,\mathbb U_{\abs b^{-\tau},\tau}$
to the Hamiltonian
\be \label{ham-D}\Hh_D:= \frac12 \A_0\B_0 +\A_+\B_++\A_-\B_-\ee
for $ab>0$ and to the Hamiltonian
\be \label{ham-C}\Hh_C:= \frac12 \A_0\B_0 + \A_-\B_++\A_+\B_-\ee
for $ab<0$ which correspond to indices $(a,\sigma), (b,\tau)$ in \eqref{bogoD} being $(1,-1),(1,1)$  or $(1,-1),(-1,1)$ respectively. 

Spectral decompositions of self-adjoint extensions of \eqref{ham-D} and \eqref{ham-C}  are essentially different thus we will discuss it separately.
In order to simplify the notation, we will replace symbols $\alpha_0(r_0,r_1)$, $\beta_0(r_0,r_1)$ by $\alpha_0$ and $\beta_0$ in all formulae below when there will be no risk of confusion.

\subsection{Spectral decomposition of $\Hh_D$}$\;$

In this case the operator $\D_0$ assumes the form
\be \D_0= \A_0+\B_0.\ee
Thus its spectrum consists of numbers $2K+\alpha_0+\beta_0$ for $K=0,1,\ldots$ and corresponding
eigenspaces  
\be \H^K_{r_0,r_1}:=\spn\{\;\ket{k}^K_{r_0,r_1}\;|\;k=0,1,\ldots,K\}\ee
are $(K+1)$-dimentional, where
\be\ket{k}^K_{r_0,r_1}:=\ket{k,K-k}_{r_0,r_1}.\ee
So we get the decomposition
\be \H_{r_0,r_1}=\bigoplus_{K=0}^\infty \H^K_{r_0,r_1}.\ee
onto subspaces invariant with respect to $\Hh_D$.
Hamiltonian $\Hh_D|_{\H^K_{r_0,r_1}}$ 
is self-adjoint and thus $\Hh_D$ admits unique self-adjoint extension in $\H\otimes\H$. Moreover it
acts in basis $\{\ket{k}^K_{r_0,r_1}\}_{k=0}^K$ as follows
\be \label{3-term-2-mode-1}\Hh_D \ket{k}^K_{r_0,r_1} = b_{k-1}\;\ket{k-1}^K_{r_0,r_1} +
a_k\;\ket k^K_{r_0,r_1}+b_k\;\ket{k+1}^K_{r_0,r_1},\ee
where
\be a_k = \frac12(2k+\alpha_0)(2(K-k)+\beta_0),\ee
\be b_k=\sqrt{(k+1)(k+\alpha_0)(K-k)(K-k+\beta_0)}. \ee
Thus, analogously to the Section \ref{sec-1mode-ham}, the problem of finding the spectral decomposition of $\Hh_D$ can be solved by means of some family of orthogonal polynomials. In this case 
the eigenvalues of  $\Hh_D|_{\H^K_{r_0,r_1}}$ are given by
\be E_n^D:=n(n+\alpha_0+\beta_0-1)+\frac12\alpha_0\beta_0\ee
for $n=0,\ldots K$ and corresponding eigenvectors assume the form
\be \ket{E_n^D}_{r_0,r_1}^K=\sum_{k=0}^K  P_k(n) \ket{k}_{r_0,r_1}^K,\ee
where
\be P_k(n):=R_k\left(E_n^D-\frac12\alpha_0\beta_0;\alpha_0-1,
\beta_0-1,K\right) =\ee
$$=\sqrt{{{\alpha_0-1+k}\choose {k}} {{\beta_0+K-k}\choose {K-k}}} \;_3F_2\left(\begin{array}{@{}c@{}|}
-k,-n,n+\alpha_0+\beta_0-1 \\
\alpha_0, -K\end{array}\;1\right)$$
are normalized Dual Hahn polynomials, see \cite{askey}. The polynomials $\{P_k\}_{k=0}^K$ form the orthonormal basis in $L^2(\R,d\omega)$ with the measure given by
\be d\omega(x)=\sum_{n=0}^K \frac{(2n+\alpha_0+\beta_0-1)(\alpha_0)_n(-K)_nK!}{(-1)^n(n+\alpha_0+\beta_0-1)_{K+1}(\beta_0)_n n!}\times\ee
$$\times\delta\left(x-E_n^D-\frac12\alpha_0\beta_0\right)dx.$$

Finally we get the following spectral decomposition
\be \Hh_D=\sum_{r_0=0}^{l_0-1}\sum_{r_1=0}^{l_1-1}\sum_{K=0}^\infty\sum_{k=0}^K E_n^D \frac{\proj{E_n^D}{E_n^D}}
{\sc{E_n^D}{E_n^D}}.\ee
In particular we see that the eigenspaces of $\Hh_D$ in $\H\otimes\H$ are infinite dimensional.

\subsection{Spectral decomposition of $\Hh_C$}$\;$

In this case the operator $\D_0$ assumes the form
\be \D_0= \A_0-\B_0.\ee
Thus its spectrum consists of numbers $2K+\alpha_0-\beta_0$ for $K\in\Z$ and corresponding
eigenspaces  
\be \H^K_{r_0,r_1}:=\spn\{\;\ket{k}^K_{r_0,r_1}\;|\;k=0,1,\ldots\}\ee
are infinite dimentional, where
\be\ket{k}^K_{r_0,r_1}:=\left\{\begin{array}{ll}
\ket{K+k,k}_{r_0,r_1} & \textrm{for } K\geq 0\\
\ket{k,k-K}_{r_0,r_1} & \textrm{for } K< 0
\end{array}\right. .\ee

So we get the decomposition
\be \H_{r_0,r_1}=\bigoplus_{K\in\Z} \H^K_{r_0,r_1}\ee
onto subspaces invariant with respect to $\Hh_C$.
The  Hamiltonian $\Hh_C|_{\H^K_{r_0,r_1}}$ also acts in basis $\{\ket{k}^K_{r_0,r_1}\}_{k=0}^\infty$ like in
previous section, i.e.:
\be \label{3-term-2-mode-2}\Hh_C \ket{k}^K_{r_0,r_1} = b_{k-1}\;\ket{k-1}^K_{r_0,r_1} +
a_k\;\ket k^K_{r_0,r_1}+b_k\;\ket{k+1}^K_{r_0,r_1},\ee
where now
\be a_k = \left\{\begin{array}{ll}
-\frac12(2(K+k)+\alpha_0)(2k+\beta_0) & \textrm{for } K\geq 0\\
-\frac12(2k+\alpha_0)(2(k-K)+\beta_0) & \textrm{for } K< 0
\end{array}\right.,\ee
\be b_k= \left\{\begin{array}{ll}
-\sqrt{(K+k+\alpha_0)(K+k+1)(k+\beta_0)(k+1)} & \textrm{for } K\geq 0\\
-\sqrt{(k+\alpha_0)(k+1)(k-K+\beta_0)(k-K+1)} & \textrm{for } K< 0
\end{array}\right..\ee
Three-term recurrence relation \eqref{1-3-term} with these coefficients is solved by normalized Continuous Dual Hahn polynomials
\be P_n(x)=S_n(-x;u,v,w):=\ee
$$:=\frac{(u+v)_n(u+w)_n}{\sqrt{\Gamma(n+u+v)\Gamma(n+u+w)\Gamma(n+v+w)n!}}\; _3F_2
\left(\begin{array}{@{}ccc|l@{}} -n & u+\sqrt x & u-\sqrt x & \\ u+v & u+w & \end{array}1\right),$$
which are orthogonal with respect to the measure
\be d\omega(x)=\abs{\frac{\Gamma(u+i\sqrt{-x})\Gamma(v+i\sqrt{-x})\Gamma(w+i\sqrt{-x})}{\Gamma(2i\sqrt{-x})}}^2 \frac{1}{2\sqrt{-x}}\theta(-x)dx+\ee
$$+\frac{\Gamma(u+v)\Gamma(u+w)\Gamma(v-u)\Gamma(w-u)}{\Gamma(-2a)}\times$$
$$\times\sum_{\substack{n=0,1,2,\ldots\\ u+n<0}}
\frac{(2u)_n(u+1)_n(u+v)_n(u+w)_n}{(u)_n(u-v+1)_n(u-w+1)_n n!}(-1)^n\delta(x-(u+n)^2)dx,$$
see \cite{askey}.
The parameters $u$, $v$, $w$ depend on the constants
$\alpha_0$, $\beta_0$ and  $K$ in the following way.
For $K\geq 0$ we have
$$u=\left\{ \begin{array}{ll}\frac12(\beta_0-\alpha_0+1) & \textrm{if }\beta_0-\alpha_0\in(-\infty,-1)\\
\frac12(\alpha_0+\beta_0-1) & \textrm{if }\beta_0-\alpha_0\in (-1,2K+1)\\
K+\frac12(\alpha_0-\beta_0+1) & \textrm{if }\beta_0-\alpha_0\in (2K+1,\infty)\end{array}\right.$$
$$v=\left\{ \begin{array}{ll}K+\frac12(\alpha_0-\beta_0+1) & \textrm{if }\beta_0-\alpha_0\in(-\infty,-1)\\
\frac12(\beta_0-\alpha_0+1) & \textrm{if }\beta_0-\alpha_0\in (-1,2K+1)\\
\frac12(\alpha_0+\beta_0-1) & \textrm{if }\beta_0-\alpha_0\in (2K+1,\infty)\end{array}\right.$$
$$w=\left\{ \begin{array}{ll}\frac12(\alpha_0+\beta_0-1) & \textrm{if }\beta_0-\alpha_0\in(-\infty,-1)\\
K+\frac12(\alpha_0-\beta_0+1) & \textrm{if }\beta_0-\alpha_0\in (-1,2K+1)\\
\frac12(\beta_0-\alpha_0+1) & \textrm{if }\beta_0-\alpha_0\in (2K+1,\infty)\end{array}\right.$$
and for $K< 0$ we have
$$u=\left\{ \begin{array}{ll}
-K+\frac12(\beta_0-\alpha_0+1) & \textrm{if }\beta_0-\alpha_0\in(-\infty,-1+2K)\\
\frac12(\alpha_0+\beta_0-1) & \textrm{if }\beta_0-\alpha_0\in (-1+2K,1)\\
\frac12(\alpha_0-\beta_0+1) & \textrm{if }\beta_0-\alpha_0\in (1,\infty)\end{array}\right.$$
$$v=\left\{ \begin{array}{ll}
\frac12(\alpha_0-\beta_0+1) & \textrm{if }\beta_0-\alpha_0\in(-\infty,-1+2K)\\
-K+\frac12(\beta_0-\alpha_0+1) & \textrm{if }\beta_0-\alpha_0\in (-1+2K,1)\\
\frac12(\alpha_0+\beta_0-1) & \textrm{if }\beta_0-\alpha_0\in (1,\infty)\end{array}\right.$$
$$w=\left\{ \begin{array}{ll}
\frac12(\alpha_0+\beta_0-1) & \textrm{if }\beta_0-\alpha_0\in(-\infty,-1+2K)\\
\frac12(\alpha_0-\beta_0+1) & \textrm{if }\beta_0-\alpha_0\in (-1+2K,1)\\
-K+\frac12(\beta_0-\alpha_0+1) & \textrm{if }\beta_0-\alpha_0\in (1,\infty).\end{array}\right. $$

Thus $\Hh_C|_{\H^K_{r_0,r_1}}$ admits unique self-adjoint extension (see \cite{akhiezer}), so $\Hh_C$ also admits unique self-adjoint extension in $\H\otimes\H$.
The operator $ F\circ \Hh_C|_{\H^K_{r_0,r_1}} \circ F^{-1}$, where $F$ is defined like in \eqref{iso-F}, acts in $L^2(\R,d\omega)$ as operator
of multiplication by $x-\frac14\big((\alpha_0-1)^2+(\beta_0-1)^2-1\big)$ and therefore the spectrum of
$\Hh_C|_{\H^K_{r_0,r_1}}$ is
\be \spec \Hh_C|_{\H^K_{r_0,r_1}}=\left(-\infty,-\frac14\big((\alpha_0-1)^2+(\beta_0-1)^2-1\big)\right)\cup \ee
$$\cup\left\{ (u+n)^2 -\frac14\big((\alpha_0-1)^2+(\beta_0-1)^2-1\big) \;|\; n=0,1,\ldots \;\wedge\; u+n<0\right\}$$
so it always consists of the continuous part (unbounded from below) and if $u<0$ there are $-[u]$ points in the discrete part.
The eigenvectors corresponding to discrete part are of the form
\be \ket{E_n^C}_{r_0,r_1}^K=\sum_{k=0}^\infty  P_k((u+n)^2) \;\ket{k}_{r_0,r_1}^K\ee
for $n=0,1,\ldots ,-[u]$.

We see that also in this case the eigenspaces of $\Hh_C$ in $\H\otimes\H$ (if they exist) are infinite dimensional.

Similarly to the one-mode case one has the coherent state representation for the Lie algebra $\mathcal A\oplus\mathcal B$ and thus also for the Hamiltonian \eqref{hamiltonian}. This representation is obtained as the tensor product of the one-mode coherent state representations. Hence all formulae concerning the two-mode coherent state representation are easily obtained by the tensoring procedure. Therefore we will not present it here. 


\section{Physical remarks}\label{cha_phys}

In the previous chapters we have found spectral decomposition of the Hamiltonians $\Hh_{\mu\nu}$ \eqref{1-ham}
and $\Hh$ \eqref{hamiltonian}, i.e. we have integrated the considered systems.

The aim of this chapter is to 
discuss the physical interpretation of the Hamiltonians $\Hh_{\mu\nu}$ and $\Hh$.
First of all let us state that we will interpret these operators as Hamiltonians of one-mode and two-mode bosonic quantum fields in non-linear medium.

We consider $\Hh$ (resp. $\Hh_{\mu\nu}$) as interaction Hamiltonians for the system and, as shown in the paper \cite{HOT},
we introduce full Hamiltonian by the formula
\be \Hh_F := \Hh_0 + e^{-i\Hh_0t}\Hh \,e^{i\Hh_0t},\ee
where $\Hh_0:= \omega_0 \n_0 + \omega_1 \n_1$ (resp. $\Hh_0:= \omega\n$) is free bosonic Hamiltonian. The solutions of Schr\"odinger equation are given by
\be \ket{\psi(t)}=e^{-i\Hh_0t}e^{-i\Hh t}\ket{\psi(0)}\ee
and allow us to compute the time evolution of the physical characteristics of the system, e.g. mean numbers of particles,
their dispersion, Fano factors, correlation functions, and squeezing factors. 

To give more explicit interpretation of $\Hh$ let us 
observe that from the formulae \eqref{ham-AB} and  \eqref{sl-operA}, \eqref{sl-operB} it follows 
that $\Hh$ is of the form:
\be \Hh=g_{00}(\n_0,\n_1)+g_{+-}(\n_0,\n_1) (\a_0^*)^{\,l_0} \a_1^{\,l_1}+g_{-0}(\n_0,\n_1) \a_0^{\,l_0}+\ee
$$+g_{0-}(\n_0,\n_1) \a_1^{\,l_1}+
 g_{--}(\n_0,\n_1) \a_0^{\,l_0} \a_1^{\,l_1}+h.c..$$
The term  $g_{00}(\n_0,\n_1)$ 
corresponds to Kerr-type effects in the non-linear medium (e.g. bistability), see 
\cite{Peng-Li}. The term $g_{+-}(\n_0,\n_1)(\a_0^*)^{\,l_0} \a_1^{\,l_1}$ 
describes intensity dependent (parametric) conversion of 
the cluster of $l_0$ bosons in the first mode into  the cluster of $l_1$ bosons in the second mode, i.e. the factor $(\a_0^*)^{\,l_0} \a_1^{\,l_1}$ describes the absorpion by the medium of one bosonic cluster and emission of another cluster, while  the factor $g_{+-}(\n_0,\n_1)$ is 
intensity dependent coupling constant.  The conjugated term describes the reverse conversion of clusters.
By analogy the remaining terms  
describe the process of intensity dependent absorption of the cluster of bosons in one or both modes. The conjugated terms describe the emission of the same clusters. The process of emission-absorption occurs with the probability depending on the functions $g_{kl}$, where $k,l=0,+,-$, 
so
it essentially depends on the number of bosons in the medium. 

For $ab>0$ the spectrum of $\Hh$ is purely discrete thus the considered system has only bound states.
On the other hand for $ab<0$ the spectrum of $\Hh$ has continuous part and (in particular cases) discrete part.
Thus the considered system has scattering states and (in some cases) bound states. Let us remark that due to the fact that in this
case the spectrum is unbounded from below, it is correct from physical point of view to consider Hamiltonian $-\Hh$.

Some representatives of integrated class of Hamiltonians, related to Examples \ref{ex-hp} and \ref{ex-2} are:
\be \Hh_{I}=\n_0+\n_1 + 2\n_0\,\n_1+\a_0^{\,2}\,\a_1^{\,2} + (\a_0^*)^2(\a_1^*)^2,\ee
\be \Hh_{II}=\n_0+\n_1 + 2\n_0\,\n_1+\a_0^{\,2}\,(\a_1^*)^2 + \a_0^{\,2}(\a_1^*)^2,\ee
\be \Hh_{III}=\n_0+\n_1 + 2\n_0\,\n_1 +\sqrt{\n_0}\; \a_0^*\,\a_1^{\,2} + \sqrt{\n_0+1}\; \a_0\,(\a_1^*)^2,\ee
\be \Hh_{IV}=\n_0+\n_1 + 2\n_0\,\n_1 +\sqrt{\n_0 \, \n_1}\; \a_0^*\,\a_1^* + \sqrt{(\n_0+1)(\n_1+1)}\; \a_0\,\a_1.\ee
Detailed physical analysis of systems described by Hamiltonians $\Hh_{III}$ and $\Hh_{IV}$ is the subject of the paper \cite{TOHJCh}.

Physical interpretation of Hamiltonians $\Hh_{\mu\nu}$ is analogous, i.e. they describe parametric absorption-emission of one-mode of bosons in non-linear medium.

Concluding we see that investigated Hamiltonians include a wide range of non-linear processes of interaction of bosons
with medium. In particular applications they can describe the models of physical systems consisting e.g. of   
photons, phonons, magnetons, or Cooper pairs.

\appendix 
\section{Spectral decomposition of one-mode Hamiltonians}
\label{ap-1-mode}

In this Appendix we present formulae for spectral decomposition of Hamiltonians \eqref{1-ham}. The problem splits into several cases.
\bigskip

\paragraph{\bf Case 1} $\nu=0,\mu\neq0$

$\Hh_{\mu\nu}$ is related to Laguerre orthonormal polynomials $P_n(x)=L_n^{(\alpha_0(r)-1)}(\frac{2x}\mu)$.
In this case
\be d\omega(x)=\frac{2}\mu\left(\frac{2x}\mu\right)^{\alpha_0(r)-1} e^{-\frac{2x}\mu}\theta\left(\frac{x}{\mu}\right)dx\ee
and spectrum $\spec(\Hh_{\mu\nu})=\R_+\cup\{0\}$ for $\mu>0$ and
$\spec(\Hh_{\mu\nu})=\R_-\cup\{0\}$ for $\mu<0$.
\bigskip

\paragraph{\bf Case 2} $\mu=0,\nu\neq0$

$\Hh_{\mu\nu}$ is also related to Laguerre orthonormal polynomials but by different formula $P_n(x)=(-1)^nL_n^{(\alpha_0(r)-1)}(\frac{2x}\nu)$.
In this case \be d\omega(x)=\frac{2}\nu\left(\frac{2x}\nu\right)^{\alpha_0(r)-1} e^{-\frac{2x}\nu}\theta\left(\frac{x}{\nu}\right)dx\ee
and spectrum $\spec(\Hh_{\mu\nu})=\R_+\cup\{0\}$ for $\nu>0$ and
$\spec(\Hh_{\mu\nu})=\R_-\cup\{0\}$ for $\nu<0$.

\bigskip

\paragraph{\bf Case 3} $\mu>0$, $\nu<0$

$\Hh_{\mu\nu}$ is related to Meixner-Pollaczek orthonormal polynomials $P_n(x)=P_n^{(\frac{\alpha_0(r)}2)}(\frac{x}{2\sqrt{-\mu\nu}};\phi)$,
for $\phi=\arccos(-\frac{\mu+\nu}{\mu-\nu})$.
In this case
\be d\omega(x)=\frac{1}{2\sqrt{-\mu\nu}}e^{(2\phi-\pi)\frac{x}{2\sqrt{-\mu\nu}}}\abs{\Gamma\left(\frac12\alpha_0(r)+i\frac{x}{2\sqrt{-\mu\nu}}\right)}^2dx\ee
and spectrum $\spec(\Hh_{\mu\nu})=\R$.

\bigskip

\paragraph{\bf Case 4} $\mu<0$, $\nu>0$

$\Hh_{\mu\nu}$ is related to Meixner-Pollaczek orthonormal polynomials
$P_n(x)=P_n^{(\frac{\alpha_0(r)}2)}(\frac{-x}{2\sqrt{-\mu\nu}};\phi)$,
for $\phi=\arccos(-\frac{\mu+\nu}{\mu-\nu})$.
In this case
\be d\omega(x)=\frac{1}{2\sqrt{-\mu\nu}}e^{(2\phi-\pi)\frac{-x}{2\sqrt{-\mu\nu}}}\abs{\Gamma\left(\frac12\alpha_0(r)-i\frac{x}{2\sqrt{-\mu\nu}}\right)}^2dx\ee
and spectrum $\spec(\Hh_{\mu\nu})=\R$.

\bigskip

\paragraph{\bf Case 5} $\mu>\nu>0$

$\Hh_{\mu\nu}$ is related to Meixner orthonormal polynomials $P_n(x)=M_n(\frac{x}{2\sqrt{\mu\nu}}-\frac{\alpha_0(r)}{2};\alpha_0(r),c)$,
for $c=\frac{\mu+\nu-2\sqrt{\mu\nu}}{\mu+\nu+2\sqrt{\mu\nu}}$.
In this case \be d\omega(x)=\sum_{n=0}^\infty \delta(x-\alpha_0(r)\sqrt{\mu\nu}-2\sqrt{\mu\nu}\;n)
\frac{(\alpha_0(r))_n}{n!}c^n\;dx\ee
and spectrum $\spec(\Hh_{\mu\nu})=\{2\sqrt{\mu\nu}\;n+\alpha_0(r)\sqrt{\mu\nu}\;|\; n=0,1,2,\ldots\}$.
Moreover the set of vectors
\be \left\{\sqrt{\frac{n!}{(\alpha_0(r))_nc^n}}\sum_{k=0}^\infty M_k(n;\alpha_0(r),c) \;\ket{k}_r \right\}_{n=0}^\infty\ee
is an orthonormal basis in $\H_r$ consisting of eigenvectors of $\Hh_{\mu\nu}$.

\bigskip

\paragraph{\bf Case 6} $\mu<\nu<0$

$\Hh_{\mu\nu}$ is related to Meixner orthonormal polynomials $P_n(x)=M_n(\frac{-x}{2\sqrt{\mu\nu}}-\frac{\alpha_0(r)}{2};\alpha_0(r),c)$,
for $c=\frac{\mu+\nu+2\sqrt{\mu\nu}}{\mu+\nu-2\sqrt{\mu\nu}}$.
In this case \be d\omega(x)=\sum_{n=0}^\infty \delta(x+\alpha_0(r)\sqrt{\mu\nu}+2\sqrt{\mu\nu}\;n)
\frac{(\alpha_0(r))_n}{n!}c^n\;dx\ee
and spectrum $\spec(\Hh_{\mu\nu})=\{-2\sqrt{\mu\nu}\;n-\alpha_0(r)\sqrt{\mu\nu}\;|\; n=0,1,2,\ldots\}$.
Moreover the set of vectors
\be \left\{\sqrt{\frac{n!}{(\alpha_0(r))_nc^n}}\sum_{k=0}^\infty M_k(n;\alpha_0(r),c) \;\ket{k}_r \right\}_{n=0}^\infty\ee
is an orthonormal basis in $\H_r$ consisting of eigenvectors of $\Hh_{\mu\nu}$.

\bigskip

\paragraph{\bf Case 7} $\nu>\mu>0$

$\Hh_{\mu\nu}$ is related to Meixner orthonormal polynomials $P_n(x)=(-1)^nM_n(\frac{x}{2\sqrt{\mu\nu}}-\frac{\alpha_0(r)}{2};\alpha_0(r),c)$,
for $c=\frac{\mu+\nu-2\sqrt{\mu\nu}}{\mu+\nu+2\sqrt{\mu\nu}}$.
In this case \be d\omega(x)=\sum_{n=0}^\infty \delta(x-\alpha_0(r)\sqrt{\mu\nu}-2\sqrt{\mu\nu}\;n)
\frac{(\alpha_0(r))_n}{n!}c^n\;dx\ee
and spectrum $\spec(\Hh_{\mu\nu})=\{2\sqrt{\mu\nu}\;n+\alpha_0(r)\sqrt{\mu\nu}\;|\; n=0,1,2,\ldots\}$.
Moreover the set of vectors
\be \left\{\sqrt{\frac{n!}{(\alpha_0(r))_nc^n}}\sum_{k=0}^\infty (-1)^kM_k(n;\alpha_0(r),c) \;\ket{k}_r \right\}_{n=0}^\infty\ee
is an orthonormal basis in $\H_r$ consisting of eigenvectors of $\Hh_{\mu\nu}$.

\bigskip

\paragraph{\bf Case 8} $\nu<\mu<0$

$\Hh_{\mu\nu}$ is related to Meixner orthonormal polynomials $P_n(x)=(-1)^nM_n(\frac{-x}{2\sqrt{\mu\nu}}-\frac{\alpha_0(r)}{2};\alpha_0(r),c)$,
for $c=\frac{\mu+\nu+2\sqrt{\mu\nu}}{\mu+\nu-2\sqrt{\mu\nu}}$.
In this case \be d\omega(x)=\sum_{n=0}^\infty \delta(x+\alpha_0(r)\sqrt{\mu\nu}+2\sqrt{\mu\nu}\;n)
\frac{(\alpha_0(r))_n}{n!}c^n\;dx\ee
and spectrum $\spec(\Hh_{\mu\nu})=\{-2\sqrt{\mu\nu}\;n-\alpha_0(r)\sqrt{\mu\nu}\;|\; n=0,1,2,\ldots\}$.
Moreover the set of vectors
\be \left\{\sqrt{\frac{n!}{(\alpha_0(r))_nc^n}}\sum_{k=0}^\infty (-1)^kM_k(n;\alpha_0(r),c) \;\ket{k}_r \right\}_{n=0}^\infty\ee
is an orthonormal basis in $\H_r$ consisting of eigenvectors of $\Hh_{\mu\nu}$.

\bigskip

\paragraph{\bf Case 9} $\mu=\nu$

In this case formula \eqref{3-term} means that $\Hh_{\mu\nu}$ is diagonal in Fock basis and  spectrum
$\spec(\Hh_{\mu\nu})=\{2\mu n + \mu\alpha_0(r) \;|\; n=0,1,2\ldots\}$.

\bigskip

\section*{Acknowledgements}
This work is partially supported by KBN grant P03A 00129.

\newcommand{\etalchar}[1]{$^{#1}$}


\begin{thebibliography}{TOH{\etalchar{+}}06}
\providecommand{\url}[1]{\texttt{#1}}
\providecommand{\urlprefix}{URL }
\expandafter\ifx\csname urlstyle\endcsname\relax
  \providecommand{\doi}[1]{doi:\discretionary{}{}{}#1}\else
  \providecommand{\doi}{doi:\discretionary{}{}{}\begingroup
  \urlstyle{rm}\Url}\fi

\bibitem[AG93]{akhiezer}
N.~I. Akhiezer, I.~M. Glazman: Theory of Linear Operators in Hilbert Space.
\newblock Dover Publications, 1993.

\bibitem[Akh65]{akhiezer-OPS}
N.~Akhiezer: The classical moment problem.
\newblock Hafner Publ. Co., 1965.

\bibitem[Bog82]{bogoliubov}
N.~N. Bogoliubov: Quantum Fields.
\newblock Benjamin-Cummings Publishing Company, 1982.

\bibitem[Chi78]{chihara}
T.~Chihara: An introduction to orthogonal polynomials.
\newblock Gordon and Breach, New York, 1978.

\bibitem[HCOT04]{HChOT}
M.~Horowski, G.~Chadzitaskos, A.~Odzijewicz, A.~Tereszkiewicz: Systems with
  intensity--dependent conversion integrable by finite orthogonal polynomials.
\newblock \emph{J. Phys. A: Math. Gen.}, \textbf{37}:6115--6128, 2004.

\bibitem[HOT02]{HOT}
M.~Horowski, A.~Odzijewicz, A.~Tereszkiewicz: Some integrable systems in
  nonlinear quantum optics.
\newblock \emph{Journal of Mathematical Physics}, \textbf{44}(2):480--506,
  2002.

\bibitem[HP40]{holstein}
T.~Holstein, H.~Primakoff: Field dependence of the intrinsic domain
  magnetization of a ferromagnet.
\newblock \emph{Phys. Rev.}, \textbf{58}, 1940.

\bibitem[JD92]{jex-drobny}
I.~Jex, G.~Drobn\'y: Quantum properties of field modes in trilinear optical
  processes.
\newblock \emph{Phys. Rev. A}, \textbf{46}(1), 1992.

\bibitem[KM91]{klein-marshalek}
A.~Klein, E.~Marshalek: Boson realizations of {L}ie algebras with applications
  to nuclear physics.
\newblock \emph{Rev. Mod. Phys.}, \textbf{63}(2), 1991.

\bibitem[KS98]{askey}
R.~Koekoek, R.~F. Swarttouw: The {A}skey-scheme of hypergeometric orthogonal
  polynomials and its $q$-analogue.
\newblock Report DU 98-17, Delft University of Technology,
  http://aw.twi.tudelft.nl/$\sim$koekoek/askey.html, 1998.

\bibitem[Odz98]{Oqspec}
A.~Odzijewicz: Quantum algebras and $q$-special functions related to coherent
  states maps of the disc.
\newblock \emph{Commun. Math. Phys.}, \textbf{192}:183--215, 1998.

\bibitem[OHT01]{OHT}
A.~Odzijewicz, M.~Horowski, A.~Tereszkiewicz: Integrable multi-boson systems
  and orthogonal polynomials.
\newblock \emph{J. Phys. A: Math. Gen.}, \textbf{34}:4353--4376, 2001.

\bibitem[OS97]{OS}
A.~Odzijewicz, M.~\'Swi\c{e}tochowski: Coherent states map for {MIC-K}epler
  system.
\newblock \emph{J. Math. Phys.}, \textbf{38}:5010--5030, 1997.

\bibitem[Per86]{perelomov}
A.~M. Perelomov: Generalized Coherent States and Their Applications.
\newblock Springer-Verlag, 1986.

\bibitem[PL98]{Peng-Li}
J.-S. Peng, G.-X. Li: Introduction to Modern Quantum Optics.
\newblock World Scientific Publishing, 1998.

\bibitem[SB84]{sukumar-buck}
C.~Sukumar, B.~Buck: Some soluble models for periodic decay and revival.
\newblock \emph{J. Phys. A: Math. Gen.}, \textbf{17}, 1984.

\bibitem[Smi61]{smirnov}
V.~Smirnov: Lectures in Higher Mathematics (in Russian).
\newblock GIFML Moscow, 1961.

\bibitem[Suk89]{sukumar}
C.~Sukumar: Revival {H}amiltonians, phase operators and non-{G}aussian squeezed
  states.
\newblock \emph{J. Modern Optics}, \textbf{36}, 1989.

\bibitem[TOH{\etalchar{+}}06]{TOHJCh}
A.~Tereszkiewicz, A.~Odzijewicz, M.~Horowski, I.~Jex, G.~Chadzitaskos:
  Explicitly solvable models of two-mode coupler in {K}err-media.
\newblock to appear, 2006.

\bibitem[VK91]{vilenkin}
N.~J. Vilenkin, A.~U. Klimyk: Representation of {L}ie Groups and Special
  Functions, volume 1: Simplest {L}ie Groups, Special Functions and Integral
  Transforms.
\newblock Kluwer Academic Publishers, 1991.

\bibitem[WT72]{walls-tindle}
D.~Walls, C.~Tindle: Nonlinear quantum effects in optics.
\newblock \emph{J. Phys. A}, \textbf{5}, 1972.

\end{thebibliography}
\end{document}